\documentclass{article}
\usepackage{emlines}

\begin{document}

\author{Vasyl Gafiychuk\thanks{%
Institute for Applied Problems of Mechanics and Mathematics, National
Academy of Sciences of Ukraine, 3b Naukova str. L'viv, 290601, Ukraine} \and %
Ihor Lubashevsky\thanks{%
Department of Physics, Moscow State University, Vavilova str., 46--92,
117333 Moscow, Russia} \and Roman Andrushkiw\thanks{%
Department of Mathematics and Center for Applied Mathematics and Statistics,
New Jersey Institute of Technology, Newark, NJ 07102, USA}}
\title{Two boundary model for freezing front propagation in biological tissue }
\date{6 August 1998 }
\maketitle

\begin{abstract}
The response of the living tissue to the effects of strong heating or
cooling can cause the blood flow rate to vary by an order of magnitude. A
mathematical model for the freezing of living tissue is formulated which
takes into account the nonlocal temperature dependence of the blood flow
rate when the temperature distribution in the tissue is substantially
nonuniform, as in cryosurgery.
\end{abstract}

INTRODUCTION

Mathematical analysis and prediction of temperature distribution in living
tissue during the process of freezing has been used in the study and
optimization of cryosurgical procedures. In the last years a number of
different approaches to describing the heat transfer process in living
tissue have been proposed (for a review see\cite{1}--\cite{10}). Within the
framework of these approaches the obtained bioheat equation is of the
\begin{equation}
c\rho \frac{\partial T}{\partial t}=\nabla (k\nabla T)-c_{b}\rho
_{b}Jf(T-T_{a})+S ~,
\end{equation}
where $T$ is the tissue temperature, $c$ and $\rho $ denote the specific
heat and density of the tissue, $c_{b}$ and $\rho _{b}$ are the specific
heat and density of blood, $J$ is the blood flow rate per unit tissue
evolume, (i.e. the volume of blood flowing through a unit volume of tissue
per unit time), $k$ is the thermal conductivity of the tissue \cite{9},\cite
{10}, $T_{a}$ is the systemic arterial blood temperature and $S$ is the
rate of metabolic heat generation. Cofactor $\ f$, ranging from 0 to 1 is
due to heat exchange between arterial and venous blood flowing through the
nearest vessels\cite{5}-\cite{5e}.

 Propagation of the freezing front $\Gamma $
is conventionally described in terms of the free boundary problem of the
Stefan-type \cite{3}:
\begin{equation}
v_{n}\rho L=-(k\nabla _{n}T)\mid _{\Gamma _{+}}+(k\nabla _{n}T)\mid _{\Gamma
_{-}} ~,
\end{equation}
\begin{equation}
T\mid _{\Gamma _{+}}=T\mid _{\Gamma _{-}}=T_{f} ~,
\end{equation}
where $L$ is the latent heat of fusion, $\Gamma _{+}$ and $\Gamma _{-}$
denote the boundaries of the freezing front on the living and frozen sides
of the tissue, respectively, and $T_{f}$ is the freezing temperature (fig.1).

A more accurate description of the heat transfer process in living tissue is
obtained if one takes into account the fact that living tissue form an
active, highly heterogeneous medium. In particular, the phase transition in
living tissue does not occur at a single temperature, but over a temperature
range. The influence of this effect has been investigated in \cite{1}. Also,
when the size of the frozen region of the tissue is small in comparison with
the characteristic length of the blood vessels that directly control the
heat exchange between the cellular tissue and blood, the heterogeneity of
living tissue has a substantial effect on the heat transfer process.
Therefore, equation (1) which models the blood flow rate in terms of a
continuous field $J(\mathbf{r})$ has to be modified.

The response of the living tissue to the effects of strong cooling or strong
heating can cause the blood flow rate to vary by an order of magnitude \cite
{6}. In general, if the temperature distribution in the tissue is
substantially nonuniform, as for example in cryosurgery, then the
temperature dependence of the blood flow rate is nonlocal and the blood flow
rate at a given point depends on certain characteristics of the temperature
distribution rate, other than the tissue temperature at the point only.
Indeed, variations in the blood flow rate at a given point are caused mainly
by the overall response of the blood collection vessels, involving arteries
of different length and radius. Hence, equation (1) should be modified to
include a relationship specifying the tissue temperature dependence on the
blood flow rate \cite{5}.

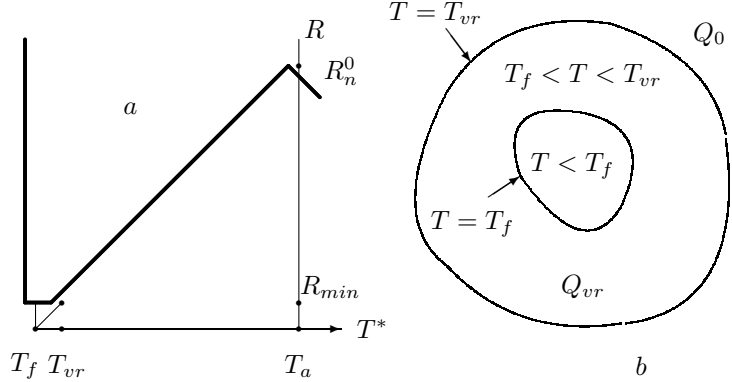
\begin{figure}[tb]
\unitlength=0.70mm
\special{em:linewidth 0.4pt}
\linethickness{0.4pt}

\begin{picture}(130.00,75.00)(-7,60)

\emline{60.00}{70.00}{1}{60.00}{125.00}{2}
\put(60.00,75.00){\circle*{1.00}}
\put(10.00,70.00){\circle*{1.00}}
\put(15.00,75.00){\circle*{1.00}}
\put(15.00,70.00){\circle*{1.00}}
\put(10.00,70.00){\vector(1,0){58.00}}
\bezier{132}(102.00,99.00)(97.00,114.00)(114.00,111.00)
\bezier{124}(114.00,111.00)(128.00,108.00)(121.00,93.00)
\bezier{136}(121.00,93.00)(114.00,82.00)(102.00,99.00)
\bezier{168}(88.00,115.00)(76.00,91.00)(88.00,82.00)
\bezier{156}(88.00,82.00)(101.00,68.00)(121.00,71.00)
\bezier{220}(122.00,71.00)(145.00,74.00)(141.00,106.00)
\bezier{140}(141.00,107.00)(138.00,122.00)(119.00,128.00)
\bezier{160}(88.00,115.00)(98.00,131.00)(119.00,128.00)
\put(60.00,120.00){\circle*{1.00}}
\put(63.00,127.00){\makebox(0,0)[cc]{$R$}}
\put(68.00,119.00){\makebox(0,0)[cc]{$R_n^0$}}
\put(66.00,78.00){\makebox(0,0)[cc]{$R_{min}$}}
\put(60.00,70.00){\circle*{1.00}}
\put(60.00,63.00){\makebox(0,0)[cc]{$T_a$}}
\put(74.00,70.00){\makebox(0,0)[cc]{$T^*$}}
\put(16.00,63.00){\makebox(0,0)[cc]{$T_{vr}$}}
\put(8.00,63.00){\makebox(0,0)[cc]{$T_f$}}
\put(28.00,112.00){\makebox(0,0)[cc]{$a$}}
\put(114.00,118.00){\makebox(0,0)[cc]{$T_f<T<T_{vr}$}}
\put(138.00,126.00){\makebox(0,0)[cc]{$Q_0$}}
\put(114.00,78.00){\makebox(0,0)[cc]{$Q_{vr}$}}
\put(112.00,101.00){\makebox(0,0)[cc]{$T<T_f$}}
\put(86.00,130.00){\makebox(0,0)[cc]{$T=T_{vr}$}}
\put(88.00,127.00){\vector(3,-4){4.67}}
\put(93.00,90.00){\makebox(0,0)[cc]{$T=T_f$}}
\put(96.00,95.00){\vector(3,2){6.00}}
\put(125.00,63.00){\makebox(0,0)[cc]{$b$}}
\emline{10.00}{75.00}{3}{10.00}{70.00}{4}
\emline{10.00}{70.00}{5}{15.00}{75.00}{6}
\special{em:linewidth 1.4pt}
\linethickness{1.4pt}
\emline{8.00}{125.00}{7}{8.00}{75.00}{8}
\emline{8.00}{75.00}{9}{13.00}{75.00}{10}
\emline{13.00}{75.00}{11}{58.00}{120.00}{12}
\emline{58.00}{120.00}{13}{64.00}{114.00}{14}
\emline{8.00}{125.00}{7}{8.00}{75.00}{8}
\emline{8.00}{75.00}{9}{13.00}{75.00}{10}
\emline{13.00}{75.00}{11}{58.00}{120.00}{12}
\emline{58.00}{120.00}{13}{64.00}{114.00}{14}

\end{picture}
\caption{
Two boundary model for tissue freezing processes: a
- the vessel resistance as a function of the blood
temperature, b - characteristic region  of living tissue.}
\label{fig27}
\end{figure}

In this paper we shall formulate a phenomenological model for the thermal
response in living tissue during the freezing process that will include the
effects of the aforementioned factors. It should be pointed out that the
main equations to be formulated can be obtained by rigorous analysis of the
microscopic relations governing the heat transfer process in living tissue
and the response of the vascular network, which we shall consider in a
subsequent paper.

TWO BOUNDARY MODEL

Let us assume that all characteristic scales of the temperature distribution
over the living tissue are greater than the mean length $l_{v}$ of the
vessels that directly control the heat exchange between the cellular tissue
and blood. In this case equation (1) is actually the result of averaging
over the scale $l_{v}$ the microscopic equation describing heat exchange
between blood in different vessels and the cellular tissue. Therefore, the
quantity $J$ appearing in equation (1) is actually the blood flow rate
averaged on the scale $l_{v}$, rather than the true blood flow rate $J_{t}$.

Nonuniformity in the true blood flow rate $J_{t}$ can be explicitly
associated with the distortion of the vascular network and, therefore, can
be characterized by scales much less than $l_v$. On scales substantially
larger than $l_{v}$ the averaged blood flow rate and the true blood flow
rate coincide, leading to the identity $J=J_{t}$. On scales less than $l_{v}$%
, the averaged blood flow rate obeys the conservation law for the total
blood flow.

A simplest relation that allows for the features mentioned above can be
expressed in the form\cite{5}-\cite{5e}
\begin{equation}
J-\nabla (\,l_{c}^{2}\,\nabla J)=J_{t} ~.
\end{equation}
The form of the second term on the left-hand side of (4) follows from the
requirement that its integral over the whole tissue domain under
consideration should be equal to zero. The mean length $l_{c}$ of the
vessels controlling the heat transfer between blood and the cellular tissue
in a given domain of radius $l_{v}$ is determined by the mean blood flow in
this domain, i.e., by the averaged blood flow rate $J$.

To find the relationship between $l_{v}$ and $J$, let us consider the heat
transfer in a given domain of diameter $l$. We assume that such a domain, on
the average, contains one artery (arteriole) and one vein (venule) of length
$l$, which is typical is the case of living tissues \cite{6}. The
characteristic rate of heat generation (or dissipation) in a domain due to
convective blood flow in these vessels can be estimated as $c_{b}\rho
_{b}(T^{\ast }-T_{a})\pi a^{2}v$, where $T^{\ast }$ and $v$ are the mean
temperature and velocity of blood in the vessels and $a$ is their radius.
Similarly, the characteristic rate of heat generation due to heat conduction
in living tissue is given approximately by $c\rho (T-T_{a})V\cdot k/(\rho
cl^{2})$, where $V\sim l^{3}$ is the volume of the domain, and $l^{2}c\rho
/k $ is the characteristic time of temperature variation on the scale $l$,
caused by heat conduction. On the scale $l_{c}$ the two quantities should be
of the same order, i.e.
\[
c\rho (T^{\ast }-T_{a})\,V_{c}\,\left( \frac{l_{c}^{2}c\rho }{k}\right)
^{-1}\sim c_{b}\rho _{b}(T-T_{a})\pi a^{2}v
\]
and the blood temperature in the corresponding vessels must be in
thermodynamic equilibrium with the cellular tissue, i.e., $T^{\ast }\sim T$.
Taking into account that $J\sim \pi a^{2}v/V_{c}$ and $c\rho \sim c_{b}\rho
_{b}$ we obtain
\begin{equation}
l_{c}^{2}\approx \frac{k}{\rho c}\cdot \frac{1}{J} ~.
\end{equation}
In particular, for typical values of $k/\rho c\sim 2\cdot 10^{-3}$ cm$^{2}$%
/s and $J\sim 3\cdot 10^{-3}$ s$^{-1}$ it follows from (5) that $l_{c}\sim 1$
cm.

Given the true blood flow rate $J_{t}$, (equation (1),(4)) and the relation
(5) determine the temperature distribution in living tissue. In order to
complete the model we must describe the response of the vascular network to
the tissue temperature variation. Since the heat exchange between cellular
tissue and blood in the vessels of length larger than $l_{c}$ is negligibly
small \cite{6}, the convective heat flow through such vessels obeys the
conservation law. Therefore, the heat flow $c_{b}\rho _{b}(T_{i}^{\ast
}-T_{a})\,I_{i}$ $(I_{i}=\pi (a^{2}v)_{i})$ in a vein $i$ of length $l>l_{c}$
is approximately equal to the total heat dissipation rate due to blood flow
in the domain $Q_{i}$, which is directly supplied with blood through this
vein and the corresponding artery. In other words, for such a vein $i$ we
have
\[
c_{b}\rho _{b}\,(T_{i}^{\ast }-T_{a})\,I_{i}=\,\int\limits_{Q_{i}}d\mathbf{r}%
\,c_{b}\rho _{b}\,J\,(T-T_{a}) ~,
\]
where the averaged blood flow rate $J$ can be replaced by the true rate $%
J_{t}$, since $l\gg l_{v}$. Upon substitution, this equation is valid
formally for veins whose length $l<l_{v}$. Indeed, in this case it reduces
to the identity
\begin{equation}
I_{i}\,=\,\int\limits_{Q_{i}}\,J_{t}d\mathbf{r}
\end{equation}
and blood in such a veins is in the thermodynamic equilibrium with cellular
tissue, i.e. $T_{i}^{\ast }\simeq T$. Thus, for any vein $i$ we may write
\begin{equation}
(T_{i}^{\ast }-T_{a})\,I_{i}\,=\,\int\limits_{Q_{i}}d\mathbf{r}%
\,J_{t}\,(T-T_{a}) ~.
\end{equation}

A valid response of the vascular network to temperature variation requires
that the resistance of each artery $i$ should depend on the mean tissue
temperature in the domain $Q_{i}$, rather than on particular details of
temperature distribution over the domain $Q_{i}$. Taking into account
equation (7), it is natural to assume that an artery $i$ responds to the
temperature $T_{i}^{*}$ of the blood in a vein $i$, because the value of $%
T_{i}^{*}$ can be treated as a mean tissue temperature in the corresponding
domain $Q_{i}$. In other words, at least in the quasistationary case, we may
suppose that the resistance $R_{i}$ to blood flow in a vein $i$ is an
explicit function of $(T_{i}^{*}-T_{a})$.

The blood flow distribution \{$I_{i}$\} over the vascular network is
directly determined by the vascular network architectonics and total
pressure drop across it, provided all the vessel resistances are have given.
Thus, when the distribution of the temperature dissipation rate $%
J_{t}\,(T-T_{a})$ in the living tissue is a given field, the explicit
dependence of the blood temperature $T_{i}^{\ast }$ in the vein $i$ on the
blood flow $I_{i}$ (through the dependence of the vessel resistances on the
blood temperature, see (7)) completely determines the blood flow pattern \{$%
I_{i}$\}. In other words, the temperature dissipation rate can be treated as
the information field for the thermoregulation of living tissue.

Hence, in the general case the evolution equation for the true blood flow
rate may be represented as
\begin{equation}
\tau \,\frac{\partial J_{t}}{\partial t}+J_{t}=J_{0}+F\{\,J_{t}\,(T-T_{a})\} ~,
\end{equation}
where the transient term allows for a possible time delay $\tau $ in the
vascular network response to the temperature variations, $J_{0}$ is a
uniform blood flow rate when $T=T_{a}$ and $F\{J_{t}(T-T_{a})\}$ is certain
functional operator of the information field $J_{t}(T-T_{a})$ that specifies
thermoregulation in the living tissue. The simplest expression for the
response operator $F$, which models thermoregulation correctly may be
written in the form
\begin{equation}
F\{J_{t}(T-T_{a})\}=\frac{1}{\Delta }\int\limits_{Q_{0}}d\mathbf{r}^{\prime
}G(\mathbf{r},\mathbf{r}^{\prime })J_{t}(\mathbf{r}^{\prime })[T(\mathbf{r}%
^{\prime })-T_{a}] ~,
\end{equation}
where $Q_{0}$ is the total living tissue domain under consideration, $G(%
\mathbf{r},\mathbf{r}^{\prime })$ is the kernel of the operator $F$, $%
\mathbf{r}$ and $\mathbf{r}^{\prime }$ are vectors, and $\Delta $ is the
length of temperature survival of the living tissue, in particular $\Delta
\approx T_{a}-T_{f}$. Typically, the temperature response of the vessels
within a tumor is strongly depressed. This allows us to set
\begin{equation}
G(\mathbf{r},\mathbf{r}^{\prime })\,=\,0 ~,
\end{equation}
where the point $\mathbf{r}^{\prime }$ belongs to the tumor domain $Q_{t}$%
,i.e. $\mathbf{r}^{\prime }\in Q_{t}$. When the tissue temperature attains a
certain low value $T_{vr}>T_{f}$, the vessels exhaust their ability to
respond to temperature variation and the blood flow rate is no longer
dependent on the tissue temperature. In mathematical terms this effect can
be taken into account if in the response operator $F\{J_{t}(T-T_{a})\}$ the
true tissue temperature $T$ is replaced by a seeming tissue temperature $%
T_{s}$, which coincides with the true tissue temperature $T$ when $T>T_{vr}$
and is equal to $T_{vr}$ for $T_{f}<T<T_{vr}$. Thus, for the given linear
response operator (9) the temperature response of the vascular network may
be described by the equation with saturation by the following equation
\begin{equation}
\tau \frac{\partial J_{t}}{\partial t}+J_{t}=J_{0}+\frac{1}{\Delta }%
\int\limits_{Q_{0}}d\mathbf{r}^{\prime }G(\mathbf{r},\mathbf{r}^{\prime
})J_{t}(\mathbf{r}^{\prime })[T_{s}(\mathbf{r}^{\prime })-T_{a}] ~,
\end{equation}
where
\begin{equation}
T_{s}=\left\{
\begin{array}{cc}
T & 
{if \ \ }T>T_{vr} \\
T_{vr} & 
{if \ }T_{f}\leq T\leq T_{vr}
\end{array}
\right. ~.
\end{equation}
In the frozen region $Q_{f}$ where $T<T_{f}$
\begin{equation}
J_{t}=0 ~.
\end{equation}
The system of equations (1),(4),(11) with boundary conditions (2),(3) and
relations (12),(13) form the desired complete phenomenological model for the
freezing process of living tissue. It should be pointed out that, within the
framework of the proposed model, the living tissue freezing process may be
represented by the propagation of two boundaries, which divide the tissue
into three regions. The first one is the frozen region, which is separated
from the extremely cooled tissue region (the second region) by the interface
$\Gamma $ where $T=T_{f}$. In the extremely cooled region the temperature
self-regulation is depressed and an effective boundary of seeming
temperature separates it from the third region, where the temperature varies
significantly in the presence of high blood flow rate.

To complete the description of our model for the freezing processes of
living tissue we need to specify the kernel $G(\mathbf{r},\mathbf{r}^{\prime
})$. The kernel should accounts for two different effect. The first of these
is increase in the blood flow rate at the point $\mathbf{r}$, associated
with additional blood flow coming from systematic circulation through the
host artery of the vascular network. The second effect is the redistribution
of the available blood flow in the vascular network. However, when the
resistance to blood flow in the vascular network is determined by a group of
vessels that very greatly in length, then the thermoregulation processes
will give rise to the formation of a low resistance path on the vascular
network that connects the host artery and the small arteries located in the
region of considerable temperature increase. In this case the blood flow
redistribution effect is ignorable in comparison with first effect mentioned
above, and the kernel $G(\mathbf{r},\mathbf{r}^{\prime })$ may be
represented approximately by the $\delta $-function
\begin{equation}
G(\mathbf{r},\mathbf{r}^{\prime })\approx \delta (\mathbf{r}-\mathbf{r}%
^{\prime })\Theta _{+}(\mathbf{r}^{\prime }) ~.
\end{equation}
where $\Theta _{+}=0$ if $\mathbf{r}\in Q_{t}$ and $\Theta _{+}=1$ for $%
\mathbf{r}\not\in Q_{t}$.

By definition, the thermal self-regulation process, where the kernel $G(%
\mathbf{r},\mathbf{r}^{\prime })$ is represented by the expression (14), is
called ideal thermoregulation.\newline

For ideal thermoregulation, equation (11) takes the following form for the
unfrozen region\cite{5}-\cite{5d}
\begin{equation}
\tau \frac{\partial J_{t}}{\partial t}+\left( \frac{T_{f}-T}{T_{f}-T_{a}}%
\right) J_{t}=J_{0} ~,
\end{equation}
when $T<T_{vr}$ and $\mathbf{r}\not\in Q_{t}$, for $\mathbf{r}\in Q_{t}$
\begin{equation}
J_{t}=J_{0} ~.
\end{equation}
In the extremely cooled region, where $T_{f}<T<T_{vr}$ and for $\mathbf{r}%
\not\in Q_{t}$
\begin{equation}
J_{t}=J_{0}\frac{T_{f}-T_{a}}{T_{f}-T_{vr}} ~.
\end{equation}
\qquad Thus, for ideal thermoregulation the description of the freezing
process for living tissue is reduced to the system of equation (1),(4),(15),
the boundary conditions (2),(3) and the expressions (14),(16),(17). It
should be noted that in this model the frozen region and the region where
the blood flow rate strongly depends on temperature are separated by a layer
(the extremely cooled region) where the blood flow rate has the maximum and
is independent of the tissue temperature.\newline

\end{document}